\documentclass[a4paper,10pt]{article}

\usepackage{amssymb}
\usepackage{amsmath}
\usepackage{graphicx}
\usepackage[dvips]{epsfig}
\usepackage[usenames,dvipsnames]{color}
\usepackage{hyperref}
\usepackage[left=.8in,right=.8in,top=.8in,bottom=.8in]{geometry}                  

\hypersetup{
    colorlinks=true,         
    linkcolor=blue,          
    citecolor=red,        
    urlcolor=Violet             
}

\newtheorem{theo}{Theorem}

\def\be{\begin{equation}}
\def\ee{\end{equation}}
\def\bea{\begin{eqnarray}}
\def\eea{\end{eqnarray}}

\title{Conformal geometrodynamics regained: gravity from duality}
\author{\bf Henrique Gomes\footnote{\href{mailto:gomes.ha@gmail.com}{gomes.ha@gmail.com}}\\\it  Department of Physics,  University of California, Davis,   CA, 95616}

\begin{document}

\maketitle

\begin{abstract}
There exist several ways of constructing general relativity from `first principles': Einstein's original  derivation,  Lovelock's results concerning the exceptional nature of the Einstein tensor from a mathematical perspective, and Hojman-Kucha\v r-Teitelboim's  derivation of the Hamiltonian form of the theory from  the symmetries of spacetime, to name a few. 
  Here I propose a different set of first principles to obtain general relativity in the canonical framework without presupposing spacetime in any way.  I  first require consistent propagation of scalar spatially covariant constraints. I find that up to a certain order in derivatives (four spatial and two temporal), there are large families of such consistently propagated constraints. Then I look for pairs of such constraints that can gauge-fix each other and form a theory with two dynamical degrees of freedom per \emph{space} point. This demand singles out the ADM Hamiltonian either in i) CMC gauge, with arbitrary (finite, non-zero) speed of light, and an extra term linear in York time, or  ii)  a gauge where the Hubble parameter is conformally harmonic. 
 \end{abstract}

\section{Introduction}

In the golden years of the canonical approach to general relativity, one of the most profound thinkers on gravity, John Wheeler, posed the famous question \cite{Wheeler}:
 ``If one did not know the Einstein-Hamilton-Jacobi equation, 
how might one hope to derive it straight off from plausible first principles, without 
ever going through the formulation of the Einstein field equations themselves?".
In response, 
Hojman, Kucha\v r and Teitelboim (HKT), in the aptly titled paper ``Geometrodynamics regained" \cite{HKT},  
tackled the problem of deriving geometrodynamics directly from  first
principles rather than by a formal rearrangement of Einstein's law. They succeeded in obtaining the canonical form of general relativity  by imposing  requirements  onto a Hamiltonian formulation ensuring that it represents a foliated space-time. 

\begin{figure}\label{figure}
 \begin{center}
 \includegraphics[width=5.0cm, height=4cm]{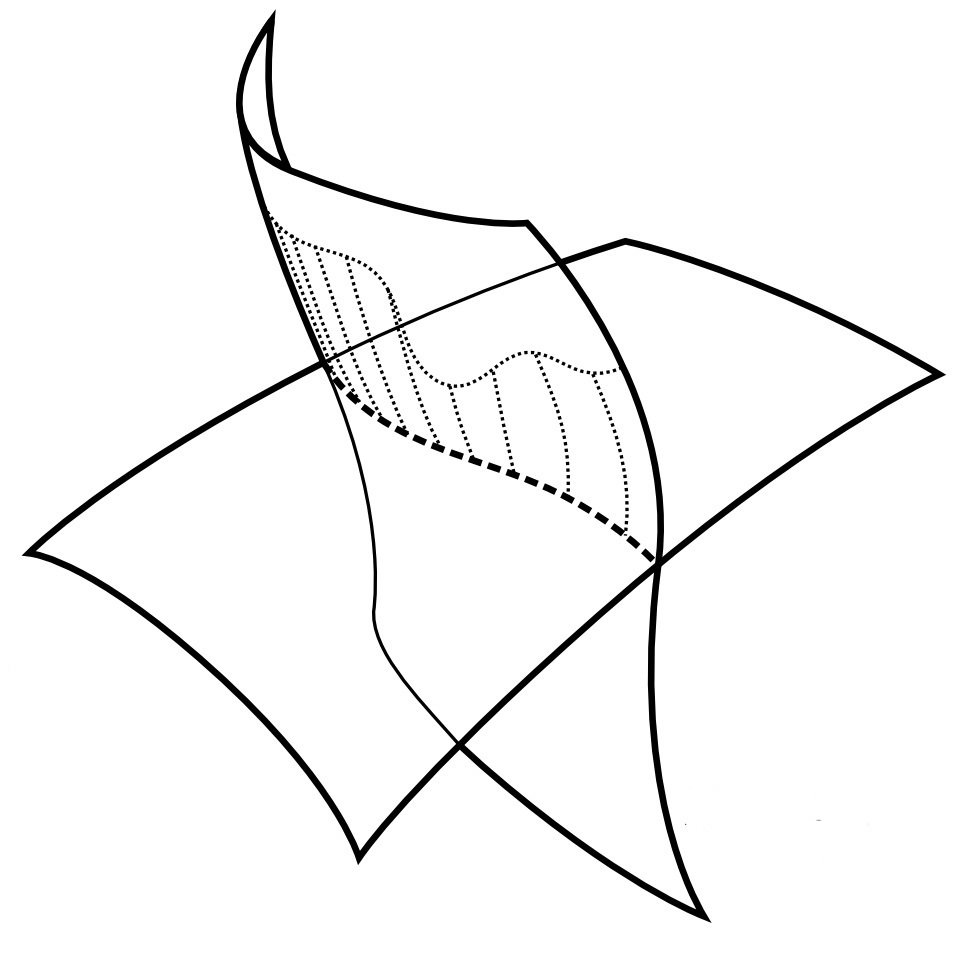}
\end{center}\caption{First-class constraint surfaces  in phase space intersecting transversally. The gauge orbits of one of the constraints is indicated by the dotted lines emanating from the intersection, where the reduced theory lies.}
\end{figure}

Here I propose a different approach to Wheeler's question. In short, I will look for what I refer to as \emph{dual symmetries} in the gravitational phase space. Dual symmetries consist of two distinct sets of constraints, which I refer to as \emph{the dual partners}. Each dual partner should be first of all a (spatially covariant) first class constraint -- which by Dirac's analysis means that each generates a (spatially covariant) symmetry -- and secondly, to fix the partnership and establish duality, one partner must gauge-fix the other. In other words, dual symmetries should be ones for which one can always find compatible obervables. 
In figure \ref{figure}, we see two first class constraint manifolds intersecting transversally, which illustrates the rather simple principle. Alternatively, this criterion amounts to searching for spatially covariant theories with two propagating degrees of freedom, which possess a gauge-fixing that is also consistentlly propagated. This dual role  arises because in the Hamiltonian formalism, a gauge-fixing condition also generates a transformation in phase space (its symplectic flow). 

The deeper reason for taking these first principles as the basis of my construction  cannot, however, be fully appreciated by only considering the classical theory. As realized in the mid 60's by Feynman, and later resolved simultaneously  by Bechi, Rouet, Stora and Tyutin \cite{BRST1, BRST2}, theories with  non-abelian symmetries require extra care upon quantization, so that pure gauge degrees of freedom don't propagate. The extended theoretical framework in which these redundancies are adequately taken into account is today known as BRST. In the Hamiltonian setting, the conditions utilized here imbue the extended BRST system with interesting properties. Namely, they ensure that with just the right gauge-fixing, the gauge-fixed,  BRST-extended Hamiltonian possesses not only the symmetries related to the original system, but also those related to its gauge-fixing. Thus the results obtained here can be said to emerge out of broad symmetry requirements, but it is surprising that we do not have to demand in advance what symmetries the emerging theory should embody, they are self selected. 

In contrast, Hojman-Kucha\v r-Teitelboim used the fact that  the set of vector fields that generate the tangential and normal deformations of an embedded hypersurface in a Riemannian manifold produce a specific vector commutation algebra, i.e. a specific symmetry. They then sought constraints  in the the space of functionals of the spatial metric $g_{ab}$ and momenta $\pi^{ab}$,  whose own commutation algebra (Poisson bracket) would mirror the hypersurface deformation algebra. Clearly, this derivation must assume the prior existence of spacetime. 
With a few other requirements, they were eventually led to the (super)momentum constraint $H_a(x)=\nabla_b\pi^{b}_{~a}(x)=0$ and the (super) Hamiltonian constraint: $S(x)=R\sqrt{g}-\frac{\pi^{ab}\pi_{ab}-\frac{1}{2}\pi^2}{\sqrt g}=0$  (where $\pi=g_{ab}\pi^{ab}$), which are responsible for the entire dynamics of space-time in  the Hamiltonian formulation of general relativity.  That is, these two functionals have associated  symplectic vector fields, which (due to their first class property)  generate symmetry transformations in phase space. The Hamiltonian (or scalar) constraint $S(x)=0$ generates (on-shell) refoliations of space-times (i.e. a description of space-time by  different sets of surfaces of siumultaneity), while the momentum constraint (or diffeomorphism constraint)  generates foliation-preserving spatial diffeomorphisms.

I cannot say if the HKT answer satisfied John Wheeler, or if any of the other construction principles that lead more or less uniquely to GR ever did. In the case of gravity there is an almost abundant variety of such principles.  Very likely the most well-known  was introduced by Lovelock \cite{Lovelock}, building on earlier work by Weyl.  Lovelock showed that the Einstein tensor is the unique second order, generally covariant, divergence-free tensor with 2 derivatives of the metric in 4-dimensions.  Other construction principles based on the massless spin-2 character of the graviton have also been introduced as early on as 1962 by Feynman \cite{Feynman_gravitation}, and as late as the thermodynamic--based derivation in 1995 by Jacobson \cite{Jacobson_Thermo}. For some of these routes, I point the interested reader to box 17.2 of \cite{MTW}, entitled ``6 Routes to Einstein's geometrodynamic law." 

 One way I can justify a shifting in the understanding of  a theory to different principles is by just appealing to the intrinsic value of alternative ontologies. Quoting HKT:  
``The importance of alternate foundations of a basic 
physical theory cannot be overexaggerated. The conceptual reformulation of a 
theory may open a new path to its development or even lead to its modification. 
Thus, Feynman's path-integral approach to quantum field theory led to the implementation of powerful approximation techniques, and Faraday's reformulation of 
action-at-a-distance stationary electrodynamics in terms of the field concept 
developed into Maxwell's electrodynamics.[...] In this spirit, believing in the potential 
fruitfulness of the canonical variational-differential approach to the general 
theory of relativity, we have undertaken the study of a Seventh Route to Einstein's."

 In the spirit of HKT, I will implement certain conditions on phase space functionals that will lead  more or less directly to my results. Unlike HKT, I will not try to match algebras of  constraints. In fact, I  will not even assume the existence of space-time (and hypersurface embeddings therein), but only of the gravitational canonical phase space. Granted, there is already some semblance of time in phase space, which is fully fleshed out when one posits a Hamiltonian. It is however still surprising that the space-time structure that we know and love, with refoliation invariance, emerges from assumptions in which it was not present. In the interest of full disclosure, departing from my first principles I will not be able to recover the full range of solutions to Einstein gravity, only those that either have constant mean curvature foliations (i.e. those space-times that can be sliced with hypesurfaces with constant trace of the extrinsic curvature), or which can be foliated by space-like surfaces for which the Hubble parameter is a conformal harmonic function, i.e. satisfies $(\nabla^2-8R)H=0$. 

The underlying reason for the restriction to CMC slicing can be seen as follows: there are constraints arising in the Hamiltonian formulation of general relativity -- also referred to as the Arnowitt-Deser-Misner (ADM) formulation \cite{ADM}--  due to a redundant description of an underlying relational theory. But it is possible to reduce the theory to a physical phase space by  the well-known York conformal method \cite{York}, which basically exhausts the spatial conformal degrees of freedom of the metric to solve the scalar initial value constraint $S(x)=0$. This construction gives rise to what is known as \emph{conformal superspace} \cite{Moncrief}. The issue is that a continuous solution curve of the reduced Hamiltonian ADM equations in conformal superspace can only uniquely rebuild space-times with complete maximal slicing or with constant trace of the momentum  (CMC).    

The special character of the trace of the momenta, $\pi$, is that it serves a double role in the formalism. It can be seen as the maximal slicing gauge fixing for ADM, $\pi=0$, but also as a generator of spatial Weyl transformations,\footnote{Also called in the mathematical literature \emph{conformal transformations}, which generates some confusion in the communication between the two fields. I should also note that for maximal slicing on compact manifolds there are restrictions on the Yamabe class of the metrics, restrictions required for the method to work.  These technical requirements can be overcome by enlarging the possible foliations to be those of \emph{constant mean curvature} (CMC). The associated conformal transformations are then those that preserve the total volume of space.  }
 \begin{eqnarray*}
 \{ \pi(\rho), g_{ab}(x)\}&=&\rho(x) g_{ab}(x)=:\delta_\rho g_{ab}(x)\\
 \{ \pi(\rho), \pi^{ab}(x)\}&=&-\rho(x) \pi^{ab}(x)=:\delta_\rho \pi^{ab}(x)
\end{eqnarray*}  
where I denoted the smearing $\int d^3x \rho \pi= \pi(\rho)$ and the canonical Poisson brackets $\{g_{ab}(x), \pi^{cd}(y)\}=\delta(x,y)\delta^{(ab)}_{(cd)}$. 
  This property, that $\pi$ generates a symmetry compatible with spatial diffeomorphisms, is at the heart of the success of the York method, and is part of what makes the entire construction of conformal superspace possible. It is the inspiration for one of the central criteria I use in my own construction principle, which can be encapsulated in the question: `Are there other sets of symmetries that gauge-fix each other, besides Weyl and refoliations?'

  In favor of the 3+1 representation, which I take as a starting point,  Dirac wrote \cite{Dirac}: ``A few decades ago it seemed quite certain that one had to express the whole of physics in four-dimensional form. But now it seems that four-dimensional symmetry is not of such overriding importance, since the description of nature sometimes gets simplified when one departs from it."   It is my duty now to show the reader why such requirements make physical and mathematical sense, and why indeed they lead to the results claimed here. I aim to show not only that one can obtain a refoliation symmetry without any assumption of an underlying spacetime,  but that the requirements are natural in a physical sense.

 It is my hope that providing this new first principles derivation of gravity (in these two very special gauges), I extend HKT's philosophy of considering the 3+1 Hamiltonian formalism not only as a technical tool, but as a self-consistent means of arriving at consistent theories.  The present results set Hamiltonian 3+1 gravity as an equivalent ontology to, and   independent from, space-time. 

\section{Construction}

My method here will be to  look for certain restrictions on  functionals in the phase space of gravity. This phase space is  coordinatized by the spatial metric and its conjugate momentum $(g_{ab}(x),\pi^{ab}(x))$ where $x\in M$, and $M$ is a 3-dimensional manifold which I assume for technical simplicity to be compact without boundary (closed). In broad terms, to be better specified later, I will first look for  scalar functionals that that are i) first class with respect to the spatial diffeomorphism constraint (also referred to as the super-momentum constraint).
 The first class property ensures that they generate symmetries themselves,  that they are consistent and  compatible with spatial diffeomorphisms.
 
 After I have a complete characterization of terms that obey this criterion, I will look for pairs that, as a pair: ii) gauge-fix each other up to a finite number of degrees of freedom, and iii) can represent a theory  with the correct number of physical dynamical propagating degrees of freedom for gravity (with the help of the spatial diffeomorphism constraint). This means that they should be scalar constraints, and, as a pair, contain the full momentum variables.
 
I will implicitly require that the constraints be regular, which ensures these constraint surfaces form submanifolds in phase space. Another requirement, which I will deem optional, is that the intersection surface between these two constraints be connected. I characterize it as optional as this requirement avoids Gribov ambiguities in the gauge-fixing, and whether one considers or not theories with Gribov ambiguities seems to me not completely homogeneous in the literature. 
 
   The second  of these requirements (ii), signifies that the generated symmetries can   be deemed dual in two (related) senses. In a simpler sense, it means that they can share observables. In the more elaborate sense it means that a BRST treatment imbues such a system with a type of supersymmetry called ``symmetry doubling" \cite{SD_Sym_doub}, which might have important properties for quantization (as does usual supersymmetry).  I will briefly discuss the necessity of requirement (ii) from this point of view at the beginning of section \ref{sec:second_class}
   
   Ultimately, one would require a single degree of freedom to remain unfixed, as is the case in shape dynamics \cite{SD_first}. The interest one might have in leaving a  single  constraint unfixed is that it  can  serve as a global reparametrization constraint in phase space. In that capacity it is equivalent to a true Hamiltonian through which variables can be said to truly evolve.   
  For a closed manifold however, the full completion of such a program requires us to also consider (mildly) non-local constraints, including also integrals of local densities constructed from the metric variables. This greatly enlarges the universe of possible constraint terms, and while the analysis is still possible, it becomes rather involved. 

  Requirement (iii) is important if we would like to interpret a remaining unfixed constraint as generating global reparametrizations in phase space, i.e. as generating dynamics. Indeed, there are many sets of constraints that are consistently propagated and which furthermore gauge fix each other at most up to a \emph{finite} number of degrees of freedom. 
  However, besides the pair `Weyl -- refoliations' and `conformal harmonic mean curvature -- refoliation' none of the other  pairs satisfying (i)-(ii) has any dependence on the traceless degrees of freedom of the momenta.  
 For these other cases, the dynamics generated by an eventual  reparametrization constraint  (leftover from either element of the dual pair) will evolve \emph{spatial geometry  only through a changing volume form}, and that makes it impossible to invert the gauge-fixed Hamiltonian  equations of motion, writing the momenta in terms of the metric, the metric velocity and Lagrange multipliers. In other words, for either member of the pair of constraints $S_1,S_2$, one obtains for $\dot g_{ab}(x)=\{S(N),g_{ab}(x)\}$ an equation of the form: 
$$\dot g_{ab}=F[g, \mbox{tr}{\pi}, N] g_{ab}
$$  
  Apart from the trace degrees of freedom, \emph{the evolution of the metric is in this case always independent from the momenta}, and whereas it can depend on the trace of the momenta, it is only insofar as  the evolution of the volume form is concerned. On the grounds that I am looking for a theory that has two \emph{dynamical}, propagating metric degrees of freedom,   I will exclude \emph{pairs} that have in them only the potential for a dynamical scale factor. 

In summary, I demand of my candidate terms the following requirements:
\begin{itemize}
\item {\bf Two dynamical metric degrees of freedom:} The constraints must be scalar. That is, they must represent one degree of freedom per space point. Furthermore, when considering possible pairs of constraints for the duality, at least one element of each  pair of constraints must include also the trace-free components of the momenta. Once more, taken in conjunction with the usual spatial diffeomorphism constraint,  this present requirement is necessary for an emerging theory with two \emph{dynamically} propagating metric degrees of freedom per space point. 
\item {\bf Consistency of symmetries and dynamics:} Each constraint  must be first class when taken in conjunction with the spatial diffeomorphism constraint (which I take to be a fundamental symmetry of my description).  
\item {\bf Symmetry duality:} I will look for  pairs of constraints where each pair has elements that are  second class  with respect to each other, up to a finite-dimensional kernel. This requirement just means that each will serve as a good gauge-fixing for the other and that observables can be related to two different symmetries. This requirement yields a type of BRST supersymmetry for the extended gauge-fixed system.   
 \item {\bf Regularity (optional):} I want my constraints to have geometrical meaning as defining manifolds in phase space, as per figure \ref{figure}. In other words, I will demand that each of the constraints define a manifold in phase space (i.e. it is regular) and that the \emph{intersection surface} be a connected manifold in phase space (so as to also avoid Gribov ambiguities).\footnote{ Technically, this implies two requirements: that the  constraints be regular functions in phase space and that the respective regular value manifold of each constraint is connected. Regularity is a very common requirement in constraint analysis, and ensures that the constraint has a geometrical meaning, locally defining a manifold in phase space. E.g. for the zero-value of a scalar function $f$, regularity is the requirement that $df_{q}\neq 0$, for all $q$ such that $f(q)=0$. The demand that the constraint surfaces be connected is required so that there are no Gribov copies from the perspective of either theory being fixed (see the last requirement in this list).} This requirement, although physically well motivated, is optional in the sense that our main results hold without it (with a Gribov copy). 
\end{itemize}
In the main part of the  paper I will consider only local terms up to four derivatives of the metric, quadratic in momenta, and with mixed terms with two spatial derivatives and one momenta.

\subsection{Setup and simplifications}

The general form of the constraints I will be considering are formed by sums, with arbitrary phase space-independent real coefficients, of arbitrary contractions between  $\nabla_a$ and $R^{ab}$ (I remind the reader that in three dimensions the Ricci curvature contains all the information of the full Riemann tensor) up to fourth order in derivatives, among the $\pi^{ab}$ up to second order, and for mixed terms of second order in spatial derivatives and first order in momenta. 
\begin{multline}\label{def:A} A(\alpha,\beta,\mu,\Lambda,j,k,l,a,b,c):= \left(\alpha R^{ab}R_{ab}+\beta \nabla^2R+\mu_n R^n+\Lambda\right)\sqrt g\\
+j\nabla^2\pi+kR^{ab}\pi_{ab} +lR\pi+\frac{a\pi^{ab}\pi_{ab}+b\pi^2}{\sqrt g}+ c\pi
\end{multline}
where $\pi=\pi^{ab}g_{ab}$, I use the summation rule for the integer parameter $n=1,2$, and used Greek letters for the coefficients corresponding to terms independent of the momenta, middle Romans $j,k$ for the mixed terms, and $a,b,c$ for the momentum-dependent ones.  All constraints that we will consider have density weight one, a choice that makes for a unified treatment of their Poisson brackets. Note also that the term $\nabla_a\nabla_b\pi^{ab}$ term does not appear due to the diffeomorphism constraint, and the contracted derivatives of the Ricci tensor don't appear, since $\nabla_a\nabla_bR^{ab}=\frac{1}{2}\nabla^2R$ by the Bianchi identity.    

I will use mainly the following tools in the computation. First, since the smearings are arbitrary, to reach a canonical form for each term resulting from the Poisson bracket, I will integrate by parts all derivatives away from one of the smearing functions. Note that, after this is done, integrating by parts is no longer helpful, as it always generates derivatives on the other smearing function. Second, 
I will then utilize an ordering in the number of derivatives of the smearing function present, since there is no crossing in this order after the previous step of establishing a canonical order. I will also assume that the full Ricci curvature and full momentum are algebraically not subsumed by the Ricci scalar and the trace of the momenta (this cannot be implied by a scalar constraint at any rate). I therefore separate the analysis to first consider the terms arising from the bracket that depend on the full form of these tensors.

A derivative ordering is also utilized for considering the constraints piecewise. In other words, since the Poisson bracket doesn't change the number of spatial derivatives, and always diminishes the momentum order by one,  terms of higher order have to close their algebra before including possible lower order terms. For instance, the Poisson bracket between $R^{ab}R_{ab}$ and $\pi^2$ will still have four spatial derivatives and one time derivative (one momentum), and thus cannot mix with the Poisson bracket between $R$ and $\pi^2$ or with that between $R^{ab}R_{ab}$  and $\pi$.

The only problem with this rationale might be that employing the weak equality one could indeed mix such terms.
 However, it is easy to note that mixing doesn't help  if,  for example 
 (after implementing commutation relations, using the Bianchi identity and the momentum constraint, etc) we cannot get rid of derivatives of the smearings contracted with  $R^{ab}$ or $\pi^{ab}$, e.g. we get  leftover  terms of the sort $\lambda\pi^{ab}\nabla_b\nabla_a\eta,~ \lambda\pi^{ab}R_a^{~c}\nabla_b\nabla_c\eta$, etc. These terms cannot be accounted for by multiples of the constraint. Since the constraint is a scalar it can only contract derivatives of its terms with derivatives of the smearing, such as in a term of the form $\lambda\nabla_c(R^{ab}R_{ab})\nabla^c\eta$ (if $R^{ab}R_{ab}$ is in the linear span of the constraints under consideration, i.e. if it is one of the terms in the scalar constraint).    It will turn out that these tools are enough. 
   
Considering the constraints in order of derivatives greatly simplifies the analysis as it allows us to approach it piecewise, but even piecewise, the calculation of the brackets, integration by parts, etc, has to be computer programmed, and I used the xAct tensor package in Mathematica. I do not include the full results in this paper, as they are quite lengthy and complicated. Instead, I give the result of the calculation and the rationale for each step.

\subsection{Consistency of symmetries and dynamics}\label{sec:first_class}
Here I will look for constraints that are first class when taken in conjunction with the diffeomorphism constraints - which I assume to be a fundamental symmetry. 
I prove the following theorem: 
 \begin{theo}\label{theo:prop}
  Given the constraints $A=0$ and $\nabla_a\pi^{ab}=0$, where $A$ is given by \eqref{def:A},  
  $A$ weakly commutes with itself for the following (non-exclusive) five families of coefficients: 
 \begin{multline}
  (\alpha,\beta,\mu_2,\mu_1,\Lambda,j,k,l,a,b,c)=
  \Big\{\mbox{I}:=(\alpha,\beta,\mu_2,\mu_1,\Lambda,0,\cdots,0),~~\mbox{II}:= (0,\cdots,\Lambda,0,0,0,a,b,c),\\ \mbox{III}:=(\alpha,3\alpha+8\mu_2,\mu_2,\mu_1,\Lambda,0,0,0,0,b,c),~~\mbox{IV}:=(0,0,0,\mu_1,\Lambda,0,0,0,a,-a/2,c),~~\mbox{V}:=(0,\cdots,j,0,-8j,0,\cdots,0)\Big\}
  \end{multline}
  \end{theo}
For instance, ADM general relativity would correspond to family IV with $\mu_1=a$ and $c=0$.   
  
 For abbreviating notation, I will denote the full combination of terms by their coefficients, it being understood that each coefficient comes with its respective term. Thus, given two smearing functions $\lambda, \eta \in C^\infty(M)$,\footnote{The space of test functions are of course more general than $C^\infty(M)$, but I will not require such an extension. } we want to find those coefficients for which: 
 \be \left\{A(\alpha,\beta,\mu_2,\mu_1,\lambda,j,k,l,a,b,c)(\lambda), A(\alpha,\beta,\mu_2,\mu_1,\lambda,j,k,l,a,b,c)(\eta)\right\}\approx 0
 \ee
where the weak equality means that it needs to vanish only when $A(\alpha,\beta,\mu_2,\mu_1,\lambda,j,k,l,a,b,c)=0$.

 \begin{itemize}
 \item {\bf Quadratic curvature and mixed terms}
 \end{itemize}
Let me start then with the highest order derivative terms. Before, let me reproduce here for convenience equation \eqref{def:A}:
\begin{multline}\nonumber A(\alpha,\beta,\mu,\lambda,j,k,l,a,b,c):= \left(\alpha R^{ab}R_{ab}+\beta \nabla^2R+\mu_n R^n+\lambda\right)\sqrt g\\
+j\nabla^2\pi+kR^{ab}\pi_{ab} +lR\pi+\frac{a\pi^{ab}\pi_{ab}+b\pi^2}{\sqrt g}+ c\pi
\end{multline}
The $A(\alpha,\beta, \mu_2)$ terms, which are those of higher derivative power, clearly commute with 
themselves, and so we can continue to lower orders, by including the mixed ones: $A(\alpha,\beta, \mu_2, j,k,l)$.  

The bracket itself is too long to be printed here (I include all the steps in the longer version of this calculation \cite{SD_construction}, and in the accompanying Mathematica file). After calculating it, integrating by parts,  and using commutation relations 
we find that there are only two terms containing both the full momentum and curvature, and contracting its indices with those of the derivatives of the smearing ($\eta$, since $\lambda$ has no more derivatives):
\be
 \lambda k^2\nabla^c\eta \left(R^{ab}\nabla_b\pi_ {ac}  +\pi^{ab} \nabla_bR _ {ac}\right)  
\ee
This implies $k=0$. The analysis goes on in this fashion,\footnote{In this paper I do not show each step for arriving at every term, for brevity.}
 until we arrive at the coefficients $\alpha=k=0, \beta=-8\mu_2, l=-8 j$ which we can't cancel based on the criteria above. These produce a bracket: 
\be\label{equ:extra_total}
2\lambda l\sqrt g  \left(\eta \nabla^2 + \nabla^a\eta \nabla_a\right) \left(-\mu_2 (R^2 - 8\nabla^2R) + 
 5 l (R\pi -8 \nabla^2\pi)\right) 
\ee
which I already wrote in a suitable form for analysis. But the form of the constraint for this choice of coefficients is:
\be\label{equ:constraint_mixed} (\mu_2 (R^2 - 8\nabla^2R)+\Lambda)\sqrt g + l(R\pi -8 \nabla^2\pi) =0
\ee
so only the choice $\mu_2=\Lambda=0$ makes it first class, since otherwise \eqref{equ:constraint_mixed} doesn't imply \eqref{equ:extra_total} vanishes. Let me call attention to the fact that the constraint $(R -8 \nabla^2)\pi=0$ is saying that the trace of the momenta (basically the Hubble parameter) is conformally harmonic, i.e. it is a harmonic function of the 3-dimensional conformal Laplacian $\Delta_C:=R -8 \nabla^2$, and it is itself conformally invariant (i.e. $\pi$ is conformally invariant).  
\begin{itemize}
\item {\bf Including the momenta.}
\end{itemize}
Now we examine the terms $A(\alpha,\beta,\mu_2,0,\cdots,0,a,b,0)$ and $A(0,0,0,j,0,-8j,0,a,b,0)$ quadratic in  momenta. 

After calculating the brackets, and using integration by parts first for the bracket of the constraint \\
$A(\alpha,\beta,\mu_2,0,\cdots,0,a,b,0)$ with itself, we obtain the terms linear in $\pi^{ab}$ and $\lambda$: 
\be\label{quad1} \lambda\pi^{ab}(-(2 \mu_2 + \alpha) \eta \nabla_b\nabla_a R + 2 \alpha \eta\nabla^c\nabla_bR_{ac} + (\beta - \alpha ) \eta\nabla^2R_{ab}\ee
using commutation relations and the Bianchi identity, we can write the term 
$$\nabla^c\nabla_bR_{ac}=\frac{\nabla_a\nabla_b R}{2}+R_a^{~d}R_{db}+RR_{ab}+R^{cd}R_{cd}g_{ab}$$
substituting this in \eqref{quad1}, after a little algebra one easily obtains that either $a=0$ or $\alpha=\beta=\mu_2=0$. 

Now, setting $a=0$, we obtain, after using the Bianchi identity, etc, the following term representing the full bracket $\{A(\alpha,\beta,\mu_2,0,\cdots,0,0,b,0)(\lambda),A(\alpha,\beta,\mu_2,0,\cdots,0,a,b,0)(\eta)\}$
\be -b\pi  \lambda \left((3 \alpha - \beta + 8 \mu_2) \eta\nabla^2R + 2 ((3 \alpha - \beta + 8 \mu_2)\nabla_b\eta\nabla^bR\right)\ee
which vanishes for the choice $3 \alpha - \beta + 8 \mu_2=0$. 
It is a trivial matter to extend this to the case including $\Lambda$, $c$ and $\mu_1$. 

Now, this exhausts the terms that have one of $\alpha,\beta, \mu_2$ different than zero. Moving on to the term  $A(0,0,0,0,j,0,-8j,0,a,b,0)$, we have that there is but one term in the bracket with itself that cannot be linearly dependent of the constraint terms: 
$$-28 j a \frac{\lambda}{\sqrt g} (\pi^{ab}  \nabla_b
\eta  \nabla_a\pi) $$
which sets $a=0$. After this is done, once again we can write the bracket as: 
\be
2 \lambda j( 2\nabla^b\eta \nabla_b+  \eta\nabla^2) (5 j (R\pi - 8 \nabla^2\pi) + 8 b\pi^2)
\ee
which, as in with equations \eqref{equ:extra_total} and \eqref{equ:constraint_mixed}, cannot be set to zero by applying the constraints unless $b=0$. The same happens with $c$.  
\begin{itemize}
\item {\bf Consistent terms with full momentum}
\end{itemize}
We move on to those that have $a\neq 0$ and $\alpha=\beta=\mu_2=j=k=l=0$, i.e. $A(0,\cdots,\mu_1,\Lambda,0,\cdots,0,a,-a/2,c)$. Straightforward calculation and integration by parts yields (after implementing the momentum constraint): 
$${2 \mu_1 (a + 2 b) \lambda (2  \nabla_b\pi  \nabla^b\eta + \eta\nabla^2\pi)}$$
whose only condition for vanishing is either $\mu_1=0$, or $b=-\frac{1}{2}a$. 
This establishes theorem \ref{theo:prop}. $\square$


\subsection{Symmetry duality}\label{sec:second_class}
As mentioned before, `symmetry duality' requires us to look for elements that are  second class  with respect to each other, up to a finite-dimensional kernel. This requirement  means that each member of the pair will serve as a good gauge-fixing for the other and that \emph{observables can be related to two different symmetries}. As I also mentioned, this requirement yields a type of BRST supersymmetry for the gauge-fixed system, a property which might be important for quantization (as it is in many field theories), and further motivates the demand for the requirement. 

\subsubsection*{The importance of the BRST extension}
In the mid 60's, when trying to calculate the 1-loop contribution to the 4-vertex gluon interaction, Feynman noticed that the resulting object did not obey the optical theorem. The issue was that one was summing over internal unphysical gluons. Longitudinal and pure gauge modes were being included in the sum, and it was difficult to excise them. The solution came in the early 70's, with the advent of the BRST extended formalism. In it, one extends the system to include the unphysical ghosts and anti-ghost particles, which `canceled' the contribution of the longitudinal and pure gauge modes. The modern approach to BRST is elegantly geometrical. In it, the physical modes correspond to the cohomology of a nilpotent differential operator $s$ representing the symmetry content of the theory. To wit, the pure gauge terms are terms in the image of $s$, whereas the gauge-invariant ones are those in the kernel of the differential $s$. In this way, one identifies the physically meaningful observables with the cohomology of $s$, $\mbox{Ker}(s)/\mbox{Im}(s)$,  which is an isomorphic space to the harmonic functions wrt $\Delta_s:=s^*s$. Using the Hodge -- de Rham orthogonal decomposition, it is easy to see that one fully accounts for summing over internal unphysical modes: 
$$\langle \Psi_1|\Psi_2\rangle= \sum_i \langle \Psi_1|\chi_i\rangle\langle \chi_i|\Psi_2\rangle= \sum_i \langle \Psi_1|\Psi_i\rangle\langle \Psi_i|\Psi_2\rangle
$$ for $\Psi_i\in  \mbox{Ker}(s)/\mbox{Im}(s)$. 

In modern times, extending the system and constructing a BRST differential is a necessary first step in quantizing gauge theories with redundant degrees of freedom. The BRST symmetries are respected by the dynamics, and in constructive approaches such as that of renormalization group flow, are demanded of theory space. 

In \cite{SD_Sym_doub}, Koslowski and I discovered an interesting property that pure constraint theories such as ADM gravity had upon a BRST extension. 
Since fully explaining the symmetry doubling property here would require the introduction of to much material not needed for the results of the paper, I leave a explanation of this `symmetry doubling' property, first uncovered in \cite{SD_Sym_doub}, to the appendix \ref{sec:appendix}. Here I merely state the result: 
\begin{theo}
Let $A$ and $B$ be two sets of  regular first class constraints that are gauge-fixings of each other, and let the system be `fully covariant',  where the Hamiltonian is made of constraints, let us say $H_o=A$, as in general relativity. 
Then the BRST gauge-fixed Hamiltonian corresponding to the BRST extension with gauge fixing fermion induced by $B$ has two BRST charges.\end{theo}

What do I mean by saying that two sets of constraints $A(x)$ and $B(x)$ gauge-fix each other? The usual criterion is that one can find a unique intersection between the orbits of one first class constraint and \emph{the} surface defined by the other.\footnote{Here I have smuggled in the assumption that each forms a single connected surface in phase space.} The infinitesimal criterion is that the bracket $\{A(x), B(y)\}\approx\Delta(x,y)$ be an invertible operator on the intersection $A(x)=B(y)=0$.

The operators $\Delta(x,y)$ can be classified in terms of their principal symbol: they can be elliptic, parabolic or hyperbolic.  Parabolic and hyperbolic operators are the ones commonly used to describe heat and wave equations, and  will immediately be disallowed by my conditions since in that case 
$$\Delta N(x):=\int d^3 x \Delta(x,y)N(x)=0$$ has an infinite-dimensional set of non-trivial solutions, for which one cannot find a prescription to invert $\Delta$ (at least not without introducing further constraints -- or gauge fixings). This breaks the duality, meaning that we cannot find observables corresponding to the two different symmetries.  For elliptic operators on a closed manifold on the other hand, there can be at most  a finite--dimensional kernel.

The second part of the proof of the present results thus goes as follows. We have the set of constraints from theorem \ref{theo:prop}, each of which forms a first class system when taken together with the diffeomorphism constraint. So that we are not wasteful, before moving on to find out which of the constraints obtained from theorem \ref{theo:prop} are dual in the sense I described above, I will implement the requirement regarding {\bf Two dynamical metric degrees of freedom}. This requires that one of the dual constraints be given  by either the family  $A(0,\cdots,\mu_1,\Lambda,0,\cdots,0,a,-a/2,c)$, or $(0,\cdots,0,\Lambda,0,\cdots,0,a,b,c)$ and $a\neq 0$ for one of the pair being considered (in which case I can always divide the constraint throughout by $a$ and still obtain the same constraint surfaces).   For the reader's convenience, I once again reproduce the possible sets from theorem \ref{theo:prop}, with this modification, and the definition \eqref{def:A} here:
\begin{multline}\nonumber A(\alpha,\beta,\mu,\lambda,j,k,l,a,b,c):= \left(\alpha R^{ab}R_{ab}+\beta \nabla^2R+\mu_n R^n+\lambda\right)\sqrt g\\
+j\nabla^2\pi+kR^{ab}\pi_{ab} +lR\pi+\frac{a\pi^{ab}\pi_{ab}+b\pi^2}{\sqrt g}+ c\pi
\end{multline}
and possible combinations:\begin{multline}\nonumber
  (\alpha,\beta,\mu_2,\mu_1,\Lambda,j,k,l,a,b,c)=
  \Big\{\mbox{I}:=(\alpha,\beta,\mu_2,\mu_1,\Lambda,0,\cdots,0),~~\mbox{II}:= (0,\cdots,\Lambda,0,0,0,a,b,c),\\ \mbox{III}:=(\alpha,3\alpha+8\mu_2,\mu_2,\mu_1,\Lambda,0,0,0,0,b,c),~~\mbox{IV}:=(0,0,0,\mu_1,\Lambda,0,0,0,a,-a/2,c),~~\mbox{V}:=(0,\cdots,j,0,-8j,0,\cdots,0)\Big\}
  \end{multline}

\begin{itemize}
\item {\bf III and IV}
\end{itemize}
 Let us start by calculating the bracket between III and IV, smeared by the functions $\lambda$ and $\eta$ (where we set $a=1$):
$$\{A(0,\cdots,\mu_1,\Lambda,0,\cdots,0,1,-1/2,c)(\eta), A(\alpha',3\alpha'+8\mu'_2,\mu'_2,\mu_1',\Lambda',0,\cdots,0,b',c')(\lambda)\}
$$
After integrating by parts to rid $\lambda$ of derivatives and using the momentum constraint, the highest order derivatives come in the term: 
$$4 (3 \alpha' + 8\mu_2')(cg^{ab}-2\pi^{ab})\nabla_b\nabla_a\nabla^2\eta$$
which, for  ellipticity requires independence of the momenta,\footnote{The momentum is a symmetric tensor, but requiring it to be strictly positive (independently of coordinate system) would require a new constraint. In other words, the principal symbol here would be $\pi^{ab}\xi_a\xi_b$ for arbitrary one forms $\xi_a$, and for it to be elliptic this contraction cannot vanish at \emph{any point}. We note that even for $R^{ab}$ there is no such obligation, as would occur for instance with certain energy condition, \emph{in space-time, and for time-like vectors}, neither of which conditions apply here.   } i.e. $3 \alpha' + 8 \mu_2'=0$, which obligates us to go to the next order of the differential operator acting on $\eta$. 

This results in a few terms, some of which cannot be undone by using the constraints:
$$-16 c\mu_2'  \sqrt gR^{ab}\nabla_a\nabla_b \eta +\pi^{ab}\frac{8}{3}(-9  \mu_1'   \nabla_a\nabla_b \eta+ \mu'_2 (3 R\nabla_b\nabla_a\eta - 8 R_a^{~c}\nabla_c\nabla_b\eta + 4 R_{ab}\nabla^2\eta))$$
Which sets $\mu_1'=\mu_2'=0$, and thus $\alpha'=0$ as well.  

\begin{itemize}
\item {\bf I and IV}
\end{itemize}
Now we move on to the bracket between I and IV:
$$\{A(0,\cdots,\mu'_1,\Lambda',0,\cdots,0,1,-1/2,c)(\eta), A(\alpha,\beta,\mu_2,\mu_1,\Lambda,0,\cdots,0)(\lambda)\}
$$
Note that $\mu_1'$ commutes through, so we can safely ignore it. 
At highest order we get again:
$$ 4\beta(\nabla_a\nabla_b\nabla^2\eta(-2\pi^{ab}+c g^{ab})
$$
which implies $\beta=0$. The next order is once again highly phase space dependent. We skip the details because they are essentially a repetition of the previous item, setting successively  $\alpha=\mu_2=\mu_1=0 $ in a straightforward manner.  

\begin{itemize}
\item {\bf I and II}
\end{itemize}
Now take the pair I and II:
$$\{A(0,\cdots,0,\Lambda',0,\cdots,0,1,b,c)(\eta), A(\alpha,\beta,\mu_2,\mu_1,\Lambda,0,\cdots,0)(\lambda)\}
$$
note that this includes as a special case, that of III and II, since the momentum terms of III will not produce and derivatives when commuted with II. The bracket gives to first order in the differential operator:
$$4\beta(-\pi^{ab}+((1+2b)\pi+c\sqrt g) g^{ab})\nabla_a\nabla_b\nabla^2\eta$$
which, again, requires $\beta=0$. It is easy to understand why this structure keeps emerging: the $\nabla^2R$ term is the only one that can contribute with 4 derivatives acting on the smearing function. The terms with quadratic curvature contribute only with two derivatives contracted/multiplied with a curvature component. Again, after setting $\beta=0$ it is straightforward to check that the second order differential operator leftover contains  terms of the form $R^{ab}\nabla_a\nabla_b\eta$,  $\pi^{ab}\nabla_a\nabla_b\eta$, $\pi^{ac}R_c^{~b}\nabla_a\nabla_b\eta$, that requires us to eventually set the remainder $\alpha=\mu_2=\mu_1=0$.

\begin{itemize}
\item {\bf IV and V}
\end{itemize}
$$\{A(0,\cdots,\mu_1,\Lambda,0,\cdots,0,1,-1/2,c)(\eta),
A(0,\cdots,j,0,-8j,0,\cdots,0)(\lambda)\}
$$
This bracket is the easiest to ascertain. Its highest order derivative operator comes out proportional to 
$$ \mu_1 j\lambda \nabla^2\nabla^2\eta
$$
which is elliptic. 
\begin{itemize}
\item {\bf II and V}
\end{itemize}
$$\{A(0,\cdots,0,\Lambda',0,\cdots,0,1,b,c)(\eta),
A(0,\cdots,j,0,-8j,0,\cdots,0)(\lambda)\}
$$
This bracket is similarly simple, giving to first order a term of the form: 
$$j \lambda\pi^{ab}\nabla_a\nabla_b\eta
$$
which sets $j=0$. 
\begin{itemize}
\item {\bf IV and IV}
\end{itemize}
As the next step, we will take the fourth member, commuted with itself (with different coefficients):
$$\{A(0,\cdots,\mu_1,\Lambda,0,\cdots,0,1,-1/2,c)(\eta), A(0,\cdots,\mu'_1,\Lambda',0,\cdots,0,a',-a'/2,c')(\lambda)\}
$$
To first order this yields: 
$$4\left( (\mu_1 c'-c'\mu_1) g^{ab} - (\mu_1'-\mu_1 a')  \pi^{ab}\right)\nabla_a\nabla_b\eta$$
And now a delicate balance must be struck, we require that 
\begin{eqnarray}\label{equ:cond1}(\mu_1'-\mu_1 a')&=0\\
\label{equ:cond2}(\mu_1 c'-c\mu'_1)&\neq 0
\end{eqnarray}. Let us investigate on what this means, at the geometric level.

 Suppose first that $a'=0$. Then $\mu_1'$ also needs to vanish by \eqref{equ:cond1}, and we fall back into a subcase of `II and IV' below. Now suppose that  $a'\neq 0$. Then we obtain the same surface by dividing the constraint throughout by $a'$ (and re-defining the arbitrary coefficients $\mu_1',c',\Lambda'$). This sets $a'=1$, which implies $\mu_1=\mu_1'$ by \eqref{equ:cond1} and $c\neq c'$ by \eqref{equ:cond2}. But now, since $a=a'$ and $\mu_1=\mu_1'$, we can identify:
 $$c\pi+\Lambda\sqrt g=c'\pi+ \Lambda'\sqrt g
 $$
 and solve for $\pi$:
 \be\label{equ:trpi} 
 \pi=\sqrt g\frac{\Lambda'-\Lambda}{c-c'}
 \ee
 Thus the system at the intersection is described by $\pi=\tau\sqrt g$, for some constant $\tau$, and: 
 \be\label{ADM_CMC}
\mu R+ \frac{\sigma^{ab}\sigma_{ab}}{\sqrt g}+(\frac{3}{8}\tau^2+c\tau+\Lambda)=0
\ee
 This is exactly the starting point for York's conformal method \cite{York}, only now with the option for a $\tau$-dependent cosmological constant. The condition for uniqueness of solutions is  that $\frac{3}{8}\tau^2+c\tau+\Lambda>0$ (note that York uses the more common $2\Lambda$, and, of course, $c=0$).

\begin{itemize}
\item {\bf II and IV}
\end{itemize}
As the last step, I take the bracket:
$$\{A(0,\cdots,\mu_1,\Lambda,0,\cdots,0,a,-a/2,c)(\eta), A(0,\cdots,0,\Lambda',0,\cdots,a',b',c')(\lambda)\}
$$
where I haven't set either $a$ or $a'$ to unity because either being non-zero would satisfy the two degrees of freedom requirement. For the highest order we obtain: 
$$8 \mu_1((c'+\pi(a'+2b')) g^{ab} - a'  \pi^{ab})\nabla_a\nabla_b\eta$$
which requires for ellipticity  $a'=0$, that $\mu_1'\neq 0$  and that the \emph{scalar} $c'+2b'\frac{\pi}{\sqrt g}$ be non zero everywhere on the intersection surface. Now we must investigate this claim, and check whether $c'=-2b'\pi$ can belong to the constraint surface IV (constraint surface II will allow any value of $\pi$). First of all, note that, by setting $a'=0$, constraint IV has become a quadratic equation, with solution:
\be\label{equ:CMC2} \pi=\frac{-c'\pm\sqrt{{c'}^2-4b'\Lambda'}}{2b'}
\ee
which can either represent two connected manifolds (and thus be ruled out by the optional {\bf Regularity}  requirement (it will yield Gribov ambiguities), or have ${c'}^2-4b'\Lambda'=0$, in which case  $c'+2b'\frac{\pi}{\sqrt g}=0$, and thus not satisfy the duality requirement. In fact, regularity of the manifold formed by $b'\frac{\pi^2}{\sqrt g}+c'\pi+\Lambda'\sqrt g=0$ also requires that the doublet:\footnote{This is obtained from the adjoint of the functional one-form $\delta II$ (this is a one-form in the phase space of gravity, in coordinates $(\delta g_{ab},\delta \pi^{ab})$ for more information regarding regularity of manifolds in the gravitational phase space see \cite{Moncrief}. )}
$$\left(g_{ab}(2b'\frac{\pi}{\sqrt g}+c'), (\Lambda'\sqrt g/2+\frac{\pi^2}{\sqrt g}/2)g^{ab}+\pi^{ab}(2b'\frac{\pi}{\sqrt g}+c')\right)
$$
be non-degenerate, which amounts to the same condition. In sum, if $b'\neq 0$, either the manifold is not connected or it is not regular, which would, according to the optional {\bf Regularity} requirement,  demand that $b'=0$ to be satisfied. This leaves us with the pair IV and II with $b'=0$. 

The requirement is optional in the sense that even if we do not care about Gribov ambiguities, and allow the regular solutions to be bona-fide dual with IV, the system one obtains \emph{at the intersection, is still described basically by ADM in CMC gauge}. In other words, the intersection would still be described as that between IV and $\pi-\tau_{\pm}=0$, where where $\tau_{\pm}$ are constants (the regular solutions to the constraint surface II when $a'=0$, given in \eqref{equ:CMC2}). The intersection requires us to write the constraint IV as:
\be\label{ADM_CMC2}
\mu R+ \frac{\sigma^{ab}\sigma_{ab}}{\sqrt g}+(\frac{3}{8}\tau_{\pm}^2+c\tau_{\pm}+2\Lambda)=0
\ee
 This is again \eqref{ADM_CMC},  only now with the option for two different times (corresponding to the two different Gribov copies).\footnote{It seems likely that by allowing higher order of momenta, one will obtain  cubic equations, and so on. This might present an obstacle for the program here since there are no closed form solutions to those. Thus in this extended context it might make sense to demand the absence of Gribov ambiguities. }

Thus I obtain the following: 
\begin{theo}\label{theo:gauge-fix1}
Given any two sets of constraints $\{A_i=0\}_{i=i_1,i_2}$ , where 
 $A_i(\alpha_i,\cdots,c_i,\Lambda_i)$ is given  by \eqref{def:A}, 
the only choices of coefficients for which: i) $A_i$ weakly commutes with itself and with the spatial diffeomorphism constraint,  ii)  $A_{i_1}$ and  $A_{i_2}$ are good gauge-fixings for each other\footnote{By `good' here I mean that their bracket defines an invertible operator (invertible Fadeev-Popov matrix) on the intersection surface. Whether one demands or not the absence of Gribov ambiguities amounts to demanding or not that the optional requirement be satisfied.} iii) the gauge-fixed system $\{A_{i_1}=0\}\cap \{A_{i_2}=0\}\cap \{\nabla_a\pi^{ab}=0\} $ has two dynamical propagating degrees of freedom,  are:
\be\label{ADM_CMCC}(\alpha,\cdots,\mu_1, a,b,c,\Lambda)=(0,\cdots,0,\mu_1,a,-a/2,c,\Lambda)~~\mbox{and}~~ (0,\cdots,0,c', \Lambda')\ee
 and
 \be\label{conformal_harmonic}
(\alpha,\cdots,\mu_1, a,b,c,\Lambda)=(0,\cdots,0,\mu_1,a,-a/2,c,\Lambda)~~\mbox{and}~~ (0,\cdots,j,0,-8j,0,\cdots,0)(\lambda)
 \ee The intersection in \eqref{ADM_CMCC} corresponds to ADM in CMC with the possibility of an added term linear in York time and a varying speed of light (for $\mu_1\neq 1$), whereas that in \eqref{conformal_harmonic} corresponds to ADM in a gauge where the Hubble parameter is a conformal harmonic function.\footnote{In many cases, such as if the metric is of positive Yamabe class, it reduces to maximal slicing gauge.  In fact, due to the discrete distribution of eigenvalues of the usual Laplacian, this gauge is also just maximal slicing for a generic metric of Yamabe type strictly negative (see discussion in conclusions). } \end{theo}

\section{Conclusions}

The requirement that a theory have a consistent gauge-fixing throughout phase space  and possess two dynamical degrees of freedom could very well be taken as first principles of a physical theory, specially when one considers quantization, where good gauge-fixings  truly become indispensable. Furthermore, the first class structure of the gauge-fixing conditions, and its compatibility with the diffeomorphisms, are some of the ingredients that make the York conformal method so powerful.  A derivation of gravitation from these first principles was essentially what I sought after, and achieved, in this paper. 
\paragraph*{Main results}
In the case of the gravitational phase space, I have shown that under certain assumptions, the conformal reduction of the canonical formulation of gravity  is one of two  theories that \emph{emerges} from these criteria. To be more precise, the theory that emerges is basically  \emph{ADM in constant mean curvature gauge, with arbitrary (finite, non-zero) speed of light and an added term linear in York time}. 
I should stress that I have not in this work investigated how different terms in these constraints can be observationally distinguished. 

The full theory is given by $\nabla_a\pi^{ab}=0$ and  equation \eqref{ADM_CMC}
 \be\nonumber
\mu R+ \frac{\sigma^{ab}\sigma_{ab}}{\sqrt g}+(\frac{3}{8}\tau^2+c\tau+\Lambda)=0
\ee 
 Einstein space-times can re-emerge from the picture,  but not all space-times, only those that have complete constant mean curvature (CMC) slicings.  Furthermore, one can always rebuild a line-element from the phase space solution curves of the reduced theory, but it sometimes happens that this line-element does not form a bona-fide geometrically 4-dimensional space-time, as the  rebuilt line element might be degenerate \cite{SD_Birkhoff}.  I have not yet begun an investigation of  solutions which possess $\mu_1\neq 1$ and $c\neq 0$, and the physical relevance of these conditions,  but it would be interesting to see if and how they can affect cosmology. For instance, the constant term being added to the cosmological constant $\tau^2+c\tau$ need no longer be monotonic, which might make for interesting "evolving cosmological constant" scenarios (strongly constrained experimentally, at least for $\mu_1=a$,  by Big Bang nucleosynthesis for example). However, it is not clear to me that these further terms indeed have any physical observability (perhaps with the addition of matter).

Constraints that are first class and also contain trace-free degrees of freedom and the metric connection (such as the one that gives rise to the dynamical part of the dual constraints and equation \eqref{ADM_CMC}) are extremely hard to come by. Their commutation contain terms that contract either the momenta or curvature tensors with derivatives of the smearings, in a way that cannot be set weakly to zero, as explained in section \ref{sec:first_class}. These properties  lead to the generalized ADM constraint, with arbitrary (finite, non-zero) speed of light, and an extra $c\pi$ term, i.e. 
$$(\mu_1 R+\Lambda)\sqrt g
+a\frac{\pi^{ab}\pi_{ab}-\frac{1}{2}\pi^2}{\sqrt g}+ c\pi$$ as the only scalar constraint included in my ansatz, that contains both the connection and the trace-free degree of freedom of the momenta.  

Similarly, the choice of the constant mean curvature constraint also has further properties  that singles it out. 
From the more foundational point of view, Julian Barbour and collaborators, motivated by relationalist ideas, came to the conclusion that physical theories should have a principle of ``relative spatial scale" \cite{Barbour:1999mf}.  From the point of view of symplectic flows in in phase space, it is quite easy to see the unique character of the diffeomorphism constraint and the $\pi=0$ constraint: they generate symmetries of the canonical variables that are independent of the base point in phase space. The fact that these two transformations act as groups (as opposed to groupoids) is what allows one to have a principal fiber bundle structure with a (semi-direct) product group structure, and thus a well-defined symmetry-reduced physical base space. It could be that these two are the only constraints that have such an action in phase space.\footnote{Another conjecture, albeit one far easier to verify.} 
\paragraph*{The conformal harmonic Hubble parameter.}
Regarding this second family of gauge-fixings - the conformal harmonic mean curvature gauge - there are a few properties that should be called attention to. First of all, it is very suggestive that the only other symmetric gauge-fixing also has such strong connections with conformal transformations.  The conformal Laplacian $\Delta_C$ is conformally covariant in the sense that a conformal transformation $\bar{g}_{ab}=\varphi^4g_{ab}$, alters the operator in the following fashion:
$\Delta_C(\varphi f)=\varphi^5\bar\Delta_C(f)$.  Note also that $\pi$ is conformally invariant in the sense that under a conformal transformation (containing also $\bar{\pi}^{ab}=\varphi^{-4}\pi^{ab}$) it is a scalar invariant, $\bar\pi(x)=\pi(x)$, thus we have a conformally covariant operator acting on a conformally invariant functional.

 Suppose that $g_{ab}$ is of positive Yamabe class \cite{Yamabe},\footnote{The proof of the Yamabe conjecture states that, given a compact closed metric manifold $(M,g)$ of dimension $\geq 3$,  there exists a conformal transformation of $g$, let us call it $\bar g$, such that $(M,\bar g)$ has constant scalar curvature. Furthermore $\bar g$ is unique up to global scaling. Here I will call the constant curvature of this conformally transformed metric $\bar{R}_g:=R_{\bar{g}}(x), ~\mbox{where}~\bar{g}=\phi_{\mbox{\tiny Y}[g]}^4 g$ and $\phi_{\mbox{\tiny Y}[g]}$ is the Yamabe conformal factor responsible for such a transformation.
}, i.e. the constant scalar curvature $\bar R$ corresponding to a unique conformally transformed metric $\bar g_{ab}$ is positive. It turns out, by min-max arguments, that then the only solution to $\Delta_C\pi=0$ on a closed manifold is the trivial solution $\pi=0$, since the kernel of an operator $\nabla^2-k^2$, for $k$ a constant,  is trivial for a closed manifold.  This means that in such cases the gauge conditions are identical to maximal slicing. Due to the discreteness of the  distribution of eigenvalues for the usual Laplacian on a compact manifold, the validity of the Yamabe conjecture, and the conformal invariance properties of the equation  $\Delta_C\pi=0$, the generic condition in phase space when the metrics are Yamabe negative are also equivalent to $\pi=0$. In fact, it is at least generically equivalent, being possibly valid everywhere. For $\bar{R}[g]=0$, one would obtain a solution of the form $\pi= c \phi^{-7}_{\mbox{\tiny Y}[g]}\sqrt{ {g}}  $, with $c$ a constant. This solution rapidly approximates $\pi=c$ for metrics with small deviations from constant scalar curvature. 

 The properties of harmonic mean curvature flows have become popular in geometry, following the success of their  cousin, the Ricci curvature flow. Even if generically equivalent to maximal slicing, it would be interesting to investigate whether a conformal harmonic mean curvature flow has any uses, in either the pure geometric setting or in the initial value formulation of gravity. These are all matters which I have not started looking into yet, and have little to say about.

\paragraph*{Including further terms}

One of the obvious extension of this work would be the inclusion of higher order terms in the ansatz \eqref{def:A}, or even the attempts at formal proofs of what kind of subsets of terms can possibly be consistently propagated. 

In this respect I would like to mention one thing that stands out among the possible extensions. In this work, I have included only terms up to second order in momenta. This is a usual restriction, which is also enforced for instance in Horava-Lifschitz or Einstein-Aether gravity. The distinction to these other formalisms is that they allow the inclusion of terms of the form $\nabla_a N$ and others operators acting on what I consider here on the non-physical Lagrange multipliers (shift and lapse). 

The inclusion of higher order momentum terms are controversial to say the least. By Ostrogradski's theorem, they can  lead to loss of unitarity, or require higher order of `momenta for momenta'. There are thus good physical, general reasons to exclude them.  This remains an issue for future work.

\subsection*{Final remarks: HKT and aesthetics}
In their paper, HKT gave their own response to Wheeler's question ``If Einstein's law is inevitable, such a
modified potential [as $R^{ab}R_{ab}$ ]  must be excluded. But what natural requirement, formulated
directly in the geometrodynamical language, does exclude it?" \begin{quote} A desire to have geometrodynamics derived from purely geometrodynamical
principles is aesthetic in its origin. There is, however, yet another motivation for
undertaking such an enterprise. The language of geometrodynamics is much closer
to the language of quantum dynamics than the original language of Einstein's
law ever was. [...] Thus, one may hope
that the first principles of geometrodynamics can be adapted more readily to the
quantum theory and lead to a deeper understanding of how the macroscopic
spacetime theory grows from the quantum geometrodynamical roots."\end{quote} What really drove the efforts of HKT was a mixture of aesthetic requirements and a hope that a purely geometrodynamic formulation would be better suited for quantization.  As my final remarks, let me take the issues of aesthetics and a  quantum mechanical appropriate representation into consideration.

Here I have not, as HKT have, taken space-time to be underlying the theory in any way, but   I see the fact that it is not based on a space-time perspective \emph{at all}, as a strength, not a weakness. Of course, interpreting the required properties that serve as the filter for the possible constraints as being the manifestation of a dual description of gravitational phenomena is  optional, and then by definition  of aesthetic character.  
However, there is also another argument, a possible mechanism for self-selection of the principles introduced here that takes this interpretation beyond the aesthetic realm.  
I thus conclude the paper with a foray into the realm of  BRST formulations of Hamiltonian field theories. 
In the case of pure constraints theories (fully covariant) - such as ADM - gauge fixing terms that are also symmetry generators possess a special role in the classical BRST formulation of the gauge-fixed theory. More specifically,   the gauge-fixed BRST Hamiltonian in this instance has the BRST symmetries related to both the original symmetry and  to that of the gauge-fixing term.
To see this simple result, termed ``symmetry doubling" \cite{SD_Sym_doub}, see the appendix below.  A sufficient condition to obtain symmetry doubling is exactly to find dual symmetries, as explained in this paper. 

Thus symmetry doubling, and the resulting supersymmetric algebra, could possibly provide a  BRST--symmetry based dynamical selection principle for the dual condition. The duality interpretation would not only be aesthetically appealing, but might provide a preferred set of theories in theory space. One could argue for example that the augmented symmetry content would make the set of theories more stable under renormalization group flows, and thus form natural candidates for universality classes \cite{Astrid, Tim_effective}, or, as suggested by Smolin, to have complementary properties in suppressing IR and UV modes in the quantization of gravity.

To close the paper, I find a citation from York very appropriate for my results \cite{York}:
``An increasing amount of evidence shows that
the true dynamical degrees of freedom of the
gravitational field can be identified directly with
the conformally invariant geometry of three-dimensional
spacelike hypersurfaces embedded in
space-time.''
I would  scratch `hypersurfaces embedded in space-time', but otherwise keep York's sentence, with what I believe to be an yet increased amount of evidence.

\appendix
\section*{Appendix}\label{sec:appendix}

 For a rank one BRST charge, related to the constraints $\chi_a$ with structure functions $U_{ab}^c$, I have
\be   \Omega= \eta^a\chi _a-\frac{1}{2}\eta^b\eta^aU_{ab}^c P_c
\ee
where $\eta^a$ are the ghosts associated to the constraint transformations, and $P_b$ the canonically conjugate ghost momenta. 

The gauge-fixed Hamiltonian is constructed by choosing a ghost number $-1$ fermion, called the gauge-fixing fermion, $\tilde\Psi=\tilde\sigma^\alpha P_\alpha+...$, where $\{\tilde\sigma^\alpha\}_{\alpha \in \mathcal A}$ is a set of proper gauge fixing conditions \cite{Gomis_antifield}. 
Denoting the BRST invariant extension of the on-shell Hamiltonian (where all constraints are set to vanish) by $H_o$, the general gauge fixed BRST-Hamiltonian is written as
\begin{equation}
 H_{\tilde\Psi}=H_o+\eta^\alpha V_\alpha^\beta P_\beta+\{\Omega,\tilde\Psi\},
\end{equation}
where $\{H_o,\chi_\alpha\}=V_\alpha^\beta\chi_\beta$ and the bracket is extended to include the conjugate ghost variables. The gauge fixing  changes the dynamics of ghosts and other non-BRST invariant functions, but maintains evolution of all BRST-invariant functions. The crux of the BRST-formalism is that the gauge-fixed Hamiltonian $H_{\tilde\Psi}$ commutes strongly with the BRST generator $\Omega$. Although  gauge symmetry is completely encoded in the BRST transformation $s  := \{\Omega, . \}$, and the gauge has been fixed, the system retains a notion of gauge-invariance through BRST symmetry.

Applying this to a generally covariant theory, i.e. a system with vanishing on-shell Hamiltonian $H_o=0$, I find that the gauge-fixed BRST-Hamiltonian takes the form
\begin{equation}
 H_{\tilde\Psi}=\{\Omega,\tilde\Psi\}.
\end{equation}
Now comes the rather simple point:  suppose that $\{\sigma^\alpha\}_{\alpha \in \mathcal A}$ is both a classical gauge fixing for $\chi_\alpha$, and also a first class set of constraints: $\{\sigma^\alpha, \sigma^\beta\}= C^{\alpha\beta}_\gamma \sigma^\gamma$.  One then can construct a nilpotent gauge-fixing $\Psi$ with the same form as the BRST charge related to the system $\{\sigma^\alpha\}_{\alpha \in \mathcal A}$,  the only difference being that ghosts and antighosts are swapped. For this I only need a gauge-fixing fermion of the form:
\be \Psi=\sigma^\alpha P_\alpha-\frac{1}{2} P_b P_aC^{ab}_c \eta^c
\ee

Using this gauge-fixing fermion would mean that the BRST extended gauge-fixed Hamiltonian would be invariant under two BRST transformations
\begin{equation}
 \begin{array}{rcl}
   s_1 . &=& \{ \Omega , . \} \\
   s_2 . &=& \{ . , \Psi \} ,
 \end{array}
\end{equation}
which follows  directly from the super-Jacobi identity and nilpotency of both $\Omega$ and $\Psi$. 
This implies that the BRST gauge-fixed Hamiltonian has an \emph{additional, dual symmetry}. In this formalism it does not distinguish between the original Hamiltonian generating evolution and the  gauge-fixing fermion, which makes them bona-fide symmetry doubling pairs. As we can see, a sufficient condition for having the symmetry doubling pairs is to find dual symmetries, as defined in the main text. 

\subsection{The resulting reduced system}

 The resulting reduced system is known by different names: the reduced Einstein Hamiltonian \cite{Moncrief}, or the 3+1 conformal reduction of ADM \cite{3+1_book}.

  To fully gauge fix the CMC constraint we introduce the following separation of variables:
\be\label{equ:new_variables}
(g_{ab},\pi^{ab})\rightarrow (|g|^{-1/3}g_{ab}, |g|,|g|^{1/3}(\pi^{ab}-\frac{1}{3}\pi g^{ab}),\frac{2\pi}{3\sqrt{ |g|}})=:(\rho_{ab},\varphi, \sigma^{ab},\tau)
\ee
To simplify matters, I choose a reference metric $\gamma_{ij}$ to determine a reference density weight.\footnote{The metric $\gamma_{ij}$ can be given by a homogeneous metric depending on the topology of the space in question. Having the reference metric as a functional of $g_{ij}$, i.e. $\gamma_{ij}[g]$, would immensely complicate the equations of motion.} 
Then I define a \emph{scalar} 
conformal factor $\phi:=|g|/|\gamma|$, and replace, in \eqref{equ:new_variables} $\varphi\rightarrow \phi$.   The variable $\tau$ will give rise to what is commonly known as \emph{York time}. The inverse transformation from the new variables to the old  is given by:
 \be\label{equ:inv_transf}\pi^{ab}=\phi^{-1/3}(\sigma^{ab}+\frac{\tau}{3}\rho^{ab})~, ~~ \mbox{and} ~~g_{ab}=\phi^{1/3}\rho_{ab}\ee

I can simultaneously do a phase space reduction by defining the variables $\phi:=\phi_o[\tau, \sigma^{ab},\rho_{ab};x)$ as a solution to the Lichnerowicz-York equation \cite{York},\footnote{Here we have taken the choice $\mu_1/b=2$ which is the usual ADM Hamiltonian.} 
\be\label{LY} \nabla^2\Omega-R\Omega+\frac{1}{8}\frac{\sigma^{ab}\sigma_{ab}}{g}\Omega^{-7}-\frac{1}{12}\tau^2\Omega^{5}= 0 
\ee
where $\Omega=e^\phi$, and by setting $\tau$ to be a spatial constant defining York time, i.e. $\dot \tau=1$. Of course, this incorporates the fact that the gauge-fixing $\tau-t=0$ is second class with respect to $S(x)=0$, and thus I must symplectically reduce to get rid of these constraints. 
I  am then  left with a genuine evolution Hamiltonian:
\be\label{equ:final_Ham} \mathcal{S}=\int dt \int _\Sigma d^3x (\sigma^{ab}\dot \rho_{ab}-\ln{\phi_o}-\sigma^{ab}\mathcal{L}_\xi \rho_{ab} )
\ee
note that the last term 
$ \sigma^{ab}\mathcal{L}_\xi \rho_{ab}$ now only generates diffeomorphisms whose flux is divergenceless (incompressible). 

\section*{Acknowledgements}
 HG was supported in part by the U.S.
Department of Energy under grant DE-FG02-91ER40674. I would like to thank Steve Carlip and Lee Smolin for reading this manuscript and giving valuable feedback. 


\end{document}